\documentclass[aps,pre,twocolumn,float]{revtex4-1}

\usepackage{xcolor,hyperref}
\hypersetup{
   colorlinks,
   linkcolor={blue!50!black},
   citecolor={blue!50!black},
   urlcolor={blue!80!black}
}

\usepackage[normalem]{ulem}
\usepackage{amsmath}
\usepackage{amssymb}
\usepackage{color}
\usepackage{hyperref}
\usepackage{graphicx}

\DeclareMathAlphabet{\mathitbf}{OML}{cmm}{b}{it}

\renewcommand{\=}{\!=\!}
\newcommand{\ket}[1]{|#1\rangle}
\newcommand{\bra}[1]{\langle #1|}
\newcommand{\braket}[2]{\langle #1|#2\rangle}
\newcommand{\fv}{\mathitbf f}
\newcommand{\uv}{\mathitbf u}
\newcommand{\xv}{\mathitbf x}

\newcommand{\yv}{\mathitbf y}

\newcommand{\zerovector}{\mathBold 0}
\renewcommand{\th}{^{\mbox{\tiny th}}}

\newcommand{\mathBold}[1]{\mbox{\boldmath$#1$}}
\newcommand{\dbar}{{\,\mathchar'26\mkern-12mu d}}
\newcommand{\partialbar}{\partial\kern-0.5em\raise0.35ex\hbox{\footnotesize /}}

\newcommand{\sFrac}[2]{{\textstyle\frac{#1}{#2}}}

\makeatother

\begin{document}
\title{Rigidity and auxeticity transitions in networks with strong bond-bending interactions}

\author{Robbie Rens and Edan Lerner}
\affiliation{Institute for Theoretical Physics, University of Amsterdam, Science Park 904, 1098 XH Amsterdam, The Netherlands}

\begin{abstract} A widely-studied model for gels or biopolymeric fibrous materials are networks with central force interactions, such as Hookean springs. Less commonly studied are materials whose mechanics are dominated by non-central force interactions such as bond-bending potentials. Inspired by recent experimental advancements in designing colloidal gels with tunable interactions, we study the micro- and macroscopic elasticity of two-dimensional planar graphs with strong bond bending potentials, in addition to weak central forces. We introduce a theoretical framework that allows us to directly investigate the limit in which the ratio of characteristic central-force to bending stiffnesses vanishes. In this limit we show that a generic isostatic point exists at $z_c\!=\!4$, coinciding with the isostatic point of frames with central force interactions in two dimensions. We further demonstrate the emergence of a stiffening transition when the coordination is increased towards the isostatic point, which shares similarities with the strain-induced stiffening transition observed in biopolymeric fibrous materials, and coincides with an auxeticity transition above which the material's Poisson's ratio approaches -1 when bond-bending interactions dominate. 
\end{abstract}
\maketitle

\section{Introduction} \label{Introduction}

In 1864 Maxwell spelled out a criterion that frames of freely-hinged struts need to satisfy in order to be mechanically stable \cite{maxwell1864calculation}: if the average connectivity $z$ is higher than a threshold value $z_c\!\equiv2\dbar$ in $\dbar$ spatial dimensions, rigidity of the frame is guaranteed, regardless of the exact way the elements are connected (as long as fluctuations in connectivity are limited, and in the absence of over-constrained clusters \cite{ellenbroek_rigidity_prl_2015}). In frames with $z\!<\! z_c$ collective modes exist that are floppy \cite{calladine1978buckminster,PhysRevLett.97.105501,during2013phonon}, which means that motion associated with these modes will respect the perfect rigidity of the struts. The gradual disappearance of such floppy modes as $z\!\to\! z_c$ is known as the jamming transition \cite{ohern2003,van_hecke,liu_review}, and has been related to various mechanical and dynamical phenomena such as the divergence of viscosity in non-Brownian suspensions \cite{suspensions} and the fragility of chalcogenide glass formers \cite{chalcogenides}. 

Most studies of jamming phenomena focus on the role of steric interactions, often modeled by some form of central forces e.g.~Hookean springs \cite{wouter_epl_2009} or hard-sphere repulsions \cite{Zamponi_hard_spheres,hard_spheres_MW}. In this work we explore the mechanical properties of a different class of materials: disordered networks in which the dominant interaction takes the form of \emph{bond bending}. Our focus is motivated by recent advancements in the fabrication of colloidal gels with tunable interactions; in particular, we were inspired by the work of Schall et al.~\cite{schall_inspiration}, who built and controlled nano- and micrometer size superstructures using critical Casimir forces on patchy colloidal particles. By measuring fluctuations of the constituent colloids, the authors of \cite{schall_inspiration} established that the stiffness of bond bending in their superstructures is much larger than the stiffness of radial interactions, by up to two orders of magnitude or more \cite{private_communication}. 

Inspired by these outstanding experiments mentioned above, we set out to address the following questions: $(i)$ what are the elastic properties of materials whose mechanics are dominated by bond-bending interactions, $(ii)$ how should the micromechanics of this class of materials be understood from a \emph{geometric} perspective, and $(iii)$ what sort of jamming phenomenology emerges in this class of systems.

In this work we consider disordered networks in two dimensions (2D) of mean coordination $z$, and introduce both radial and bond bending interactions, characterized by stiffnesses $k_r$ and $k_\theta$, respectively. We consider the ratio $\mu \!\equiv\! k_r/k_\theta$ between these stiffnesses as a key tunable parameter of the material (in addition to the coordination $z$), and investigate numerically and theoretically the behavior of elastic moduli under variations of $z$ and $\mu$. We show that as the limit $\mu\!\to\!0$ is approached, scaling behavior of elastic moduli emerges, as a function of the distance between the mean connectivity and the system's jamming point, shown in what follows to coincide with the Maxwell threshold \cite{maxwell1864calculation} $z_c\!=\!4$ at which the generic, `central-force' isostatic point occurs. 

We take two complementary routes in order to study theoretically the limit $\mu\!\to\!0$ at which the scaling behavior of elastic moduli emerges. First, we fix $k_\theta$ and set $k_r\!=\!0$; this arrangement allows us to understand the scaling behavior of elastic moduli in the hyperstatic regime $z\!>\!z_c$. We further put forward a scaling argument supported by numerical tests that the hyperstatic regime is characterized by a diverging length $\ell_c\!\sim\!(z-z_c)^{-1/2}$. 

Even more intruiging is the limit $\mu\!\to\!0$ obtained by fixing $k_r$ and sending $k_\theta\!\to\!\infty$; this is achieved by considering the angles formed between bonds that share a common node to be entirely fixed, i.e.~they serve as \emph{geometric} constraints. In this limit, and away from the isostatic point, hypostatic networks feature elastic moduli $\sim k_r$. Interestly, we find that the shear modulus $G$ \emph{diverges} as $z$ is made to approach $z_c$ from below, while the bulk modulus $K$ remains regular. We present a theoretical framework that allows us to predict the divergence of $G$ with $z_c\!-\! z$ in this limit, and find good agreement with our numerical calculations. 

Finally, we consider the auxeticity of our model material; we find that far into the hypostatic regime the material features a positive Poisson's ratio, of around $0.3$, a value characteristic to many disordered materials \cite{metallic_glasses_poissons_ratio}. However, $\nu$ rapidly decreases as $z$ is increased. Interestingly, in the limit $\mu\!\to\!0$ the material approaches perfect auxeticity $\nu\!\to\!-1$ as $z\!\to\! z_c$, and remains perfectly auxetic in the entire hyperstatic regime. 

Our work is structured as follows; in Sect.~\ref{model_and_observables} we spell out the model ingredients and observables considered in our study. In Sect.~\ref{numerics} we present a numerical investigation of the elastic properties of our model, as a function of the two key control parameters, namely the ratio of stiffnesses $\mu\!\equiv\!k_r/k_\theta$ and the mean coordination $z$. In Sect.~\ref{isostatic_point} we explain the occurrence of an isostatic point at $z_c\!=\!4$ in our model. In Sect.~\ref{hyperstatic} we consider the hyperstatic regime, and study theoretically the limit $\mu\!\to\!0$, while Sect.~\ref{hypostatic} presents a theoretical framework that allows us to study the hypostatic regime in the limit $\mu\!\to\!0$. In Sect.~\ref{lengthscale} we provide scaling arguments and  show numerically that a characteristic lengthscale diverges as the isostatic point is approached. Sect.~\ref{auxeticity} discusses the auxeticity of our model, and our work is summarized in Sect.~\ref{discussion}, where we discuss future research directions.

\section{Model and key observables}
\label{model_and_observables}
\begin{figure}[!ht]
\includegraphics[width= 0.48\textwidth]{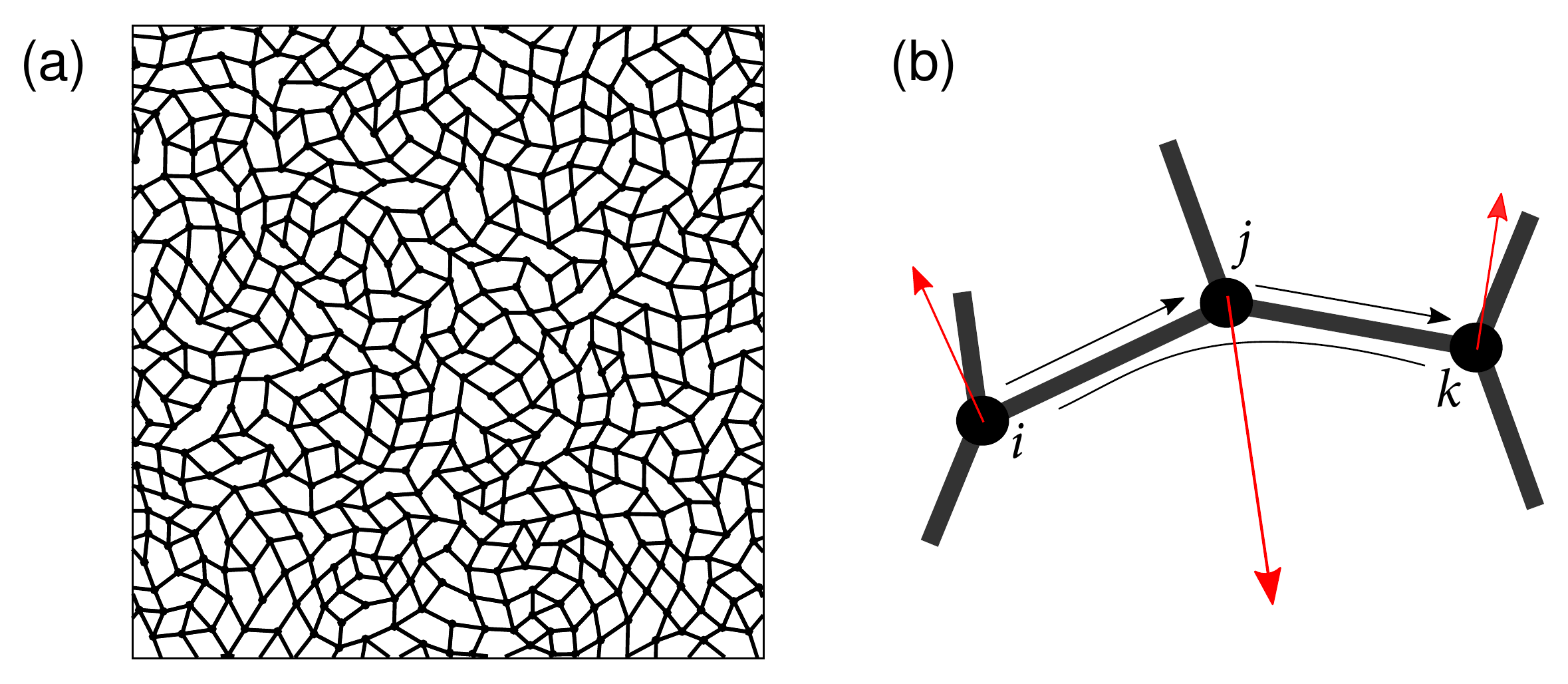}
\caption{ (a) An example of a typical network with a mean coordination of $z=3.95$ considered in this work. Our network generation protocol is described in Appendix~\ref{network_generation_protocol}. (b) Bond-bending interactions are defined on pairs of edges that connect to a common central node. The red arrows represent the field $\frac{\partial\theta_{ijk}}{\partial\xv_\ell}$, see text for details.}
\label{fig1}
\end{figure}

We consider 2D disordered planar graphs (networks) with periodic boundary considitions, whose topology is characterized by the mean number of edges per node, denoted by $z$. We build these disordered networks by adopting the contact network of dense packings of soft spheres, and pruning edges according to a protocol that maintains low node-to-node fluctuations of connectivity. Our network generation protocol is described in Appendix~\ref{network_generation_protocol}, and an example of a network with $z\!=\!3.95$ is shown in Fig.~\ref{fig1}a.

We introduce the following potential energy $U$ for our disordered networks
\begin{equation}\label{potential_energy}
U = \frac{k_\theta \bar{\ell}^2}{2}\sum_{\left< i,j,k\right>} \Delta{ \theta_{ijk} }^2 + \frac{k_r}{2}\sum_{\left< i,j\right>} \Delta{ r_{ij} }^2 \,,
\end{equation}
where $k_\theta$ and $k_r$ denote the bond-bending and central-force stiffnesses, respectively, and $\bar{\ell}$ denotes the microscopic units of length. The first term on the RHS of Eq.~(\ref{potential_energy}) represents a sum over all angles $\theta_{ijk}$ formed between pairs of edges that share a common node, with no other edges found in between the said pair, as illustrated in Fig.~\ref{fig1}b. We define the deviations from the rest-angles $\Delta\theta_{ijk}\!\equiv\!\theta_{ijk}\! -\! \theta_{ijk}^{(0)}$ with $\theta_{ijk}^{(0)}$ denoting the initial (ground state) rest-angles. The second term on the RHS of Eq.~(\ref{potential_energy}) represents a sum over the network's edges, and we define the deviation from the rest-lengths $\Delta r_{ij}\!\equiv\! r_{ij}\!-\!\ell_{ij}$ where $\ell_{ij}$ denotes the rest-length of the edge connecting the $i,j$ pair of nodes. Lengths are expressed in terms of $\bar{\ell}$ which denotes the mean rest-length. Below we will always assume that the networks reside at their respective ground states, i.e.~all of the angles are equal to their rest-angles, and all edges reside at their rest-lengths, implying that $U\!=\!0$ and that there are no stresses in the material. 

In what follows we consider simple shear and expansive strains that result from the application of the affine transformation $H(\gamma,\eta)$ on coordinates $\xv$ as $\xv\!\to\!H\!\cdot\!\xv$. The transformation $H(\gamma,\eta)$ is most conveniently parameterized by simple shear and expansive strain parameters $\gamma$ and $\eta$, respectively, and has the following form in 2D
\begin{equation}
H = \left( \begin{array}{cc}1+\eta&\gamma\\0&1+\eta\end{array}\right)\,.
\end{equation}
Using this transformation, the strain tensor $\epsilon$ assumes the form
\begin{equation}
\epsilon(\gamma,\eta) = \frac{1}{2}\!\left( H^T\cdot H - {\cal I}\right) = \frac{1}{2}\!\left( \begin{array}{cc}2\eta\!+\! \eta^2&\gamma\! +\! \gamma\eta\\\gamma \!+\! \gamma\eta&2\eta\! +\! \eta^2\! +\!\gamma^2\end{array}\right)\,,
\end{equation}
where ${\cal I}$ represents the identity tensor. We note that quadratic terms in the strain parameters $\eta$ and $\gamma$ are kept in the entries of $\epsilon$ since we will be interested in second-order derivatives with respect to those parameters, as shown below. Given a strain tensor that describes an imposed deformation mode, distances $r_{ij}\=\sqrt{\xv_{ij}\cdot\xv_{ij}}$ between the coordinates of any two nodes $\xv_i,\xv_j$ vary under such imposed deformations according to 
\begin{equation}
\delta r_{ij} \simeq  \frac{\xv_{ij}\cdot\epsilon\cdot\xv_{ij}}{r_{ij}} - \frac{1}{2} \frac{(\xv_{ij}\cdot\epsilon\cdot\xv_{ij})^2}{r_{ij}^3}\,,
\end{equation}
where $\xv_{ij}\!\equiv\!\xv_j\!-\!\xv_i$. We note importantly that in what follows we only consider either simple shear or expansive strains, and not combinations of these. 

\begin{figure*}[htbp]
\includegraphics[width= 0.9\textwidth]{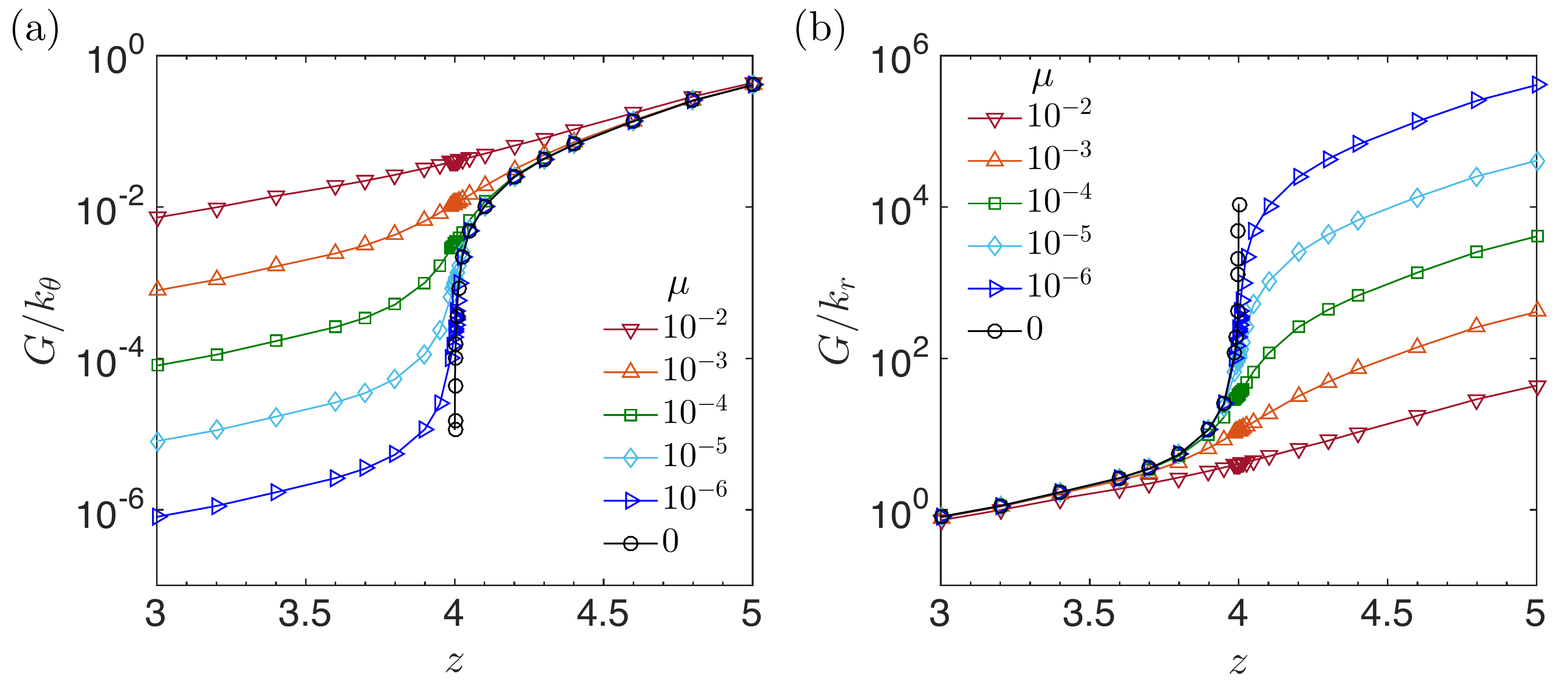}
\caption{Shear modulus $G$, rescaled by (a) $k_\theta$, and (b) $k_r$, vs.~the mean coordination $z$ for various values of $\mu\!\equiv\!k_r/k_\theta$. In this representation we see that $(i)$ a jamming transition occurs at $z_c\!=\!4$, $(ii)$ that $G\!\sim\!k_\theta$ deep in the hyperstatic regime $z\!>\! z_c$, and $(iii)$ that $G\!\sim\!k_r$ deep in the hypostatic regime $z\!<\! z_c$. The black circles correspond to the limit $\mu\!\to\!0$, discussed in detail separately for the hyperstatic regime in Sect.~\ref{hyperstatic} and for the hypostatic regime in Sect.~\ref{hypostatic}.}
\label{shear_modulus_vs_coordination}
\end{figure*}

We focus on macroscopic elastic properties as seen in the athermal shear and bulk moduli, denoted as $G$ and $K$, respectively. The athermal shear modulus is defined as
\begin{equation}\label{shearmodulus}
G \equiv  \frac{1}{V}\frac{d^2U}{d\gamma^2}\,,
\end{equation}
whereas the athermal bulk modulus is given by
\begin{equation}
K \equiv \frac{1}{V}\left(\frac{1}{4}\frac{d^2U}{d\eta^2} - \frac{1}{2}\frac{dU}{d\eta}\right)\,.
\end{equation}
Total dervatives e.g.~$d/d\gamma$ are understood as taken in the athermal limit, i.e.~under the constraints dictated by mechanical equilibrium \cite{lutsko}.

\section{Elastic properties of bond-bending-dominated networks}
\label{numerics}
We start the presentation of our results with a numerical investigation of the shear modulus variation under changes of the mean coordination $z$ and the ratio $\mu$ of bond-bending to central-force stiffnesses.  In Fig.~\ref{shear_modulus_vs_coordination} we plot the sample-to-sample means of the shear modulus averaged over 20 random networks of $N\!=25600$ nodes. The left panel plots the ratio $G/k_\theta$ vs.~the coordination $z$; noticeably, $z\!=\!4$ marks the onset of an underlying jamming transition, that becomes more pronounced as $\mu\!\to\!0$. For small $\mu$ the ratio $G/k_\theta$ grows by several orders of magnitude as $z$ approaches the critical coordination $z_c\!=\!4$, in a fashion reminiscent of the strain-stiffening transition observed upon deformation of biopolymeric fibrous materials \cite{Broedersz2011,rens2016nonlinear,jansen2018,robbies_pre,Merkel6560}. Far above $z_c$ the ratio $G/k_\theta$ becomes roughly independent of $\mu$, indicating that in this regime $G\!\sim\!k_\theta$. 

In the right panel of Fig.~\ref{shear_modulus_vs_coordination} we show the same sample-to-sample means of the shear modulus, this time rescaled by $k_r$, to find that deep in the hypostatic regime $z\!<\!4$, $G\!\sim k_r$. The data pertaining to $\mu\!=\!0$ in Fig.~\ref{shear_modulus_vs_coordination} will be discussed in detail in what follows. 

\begin{figure}[h]
\includegraphics[width= 0.48\textwidth]{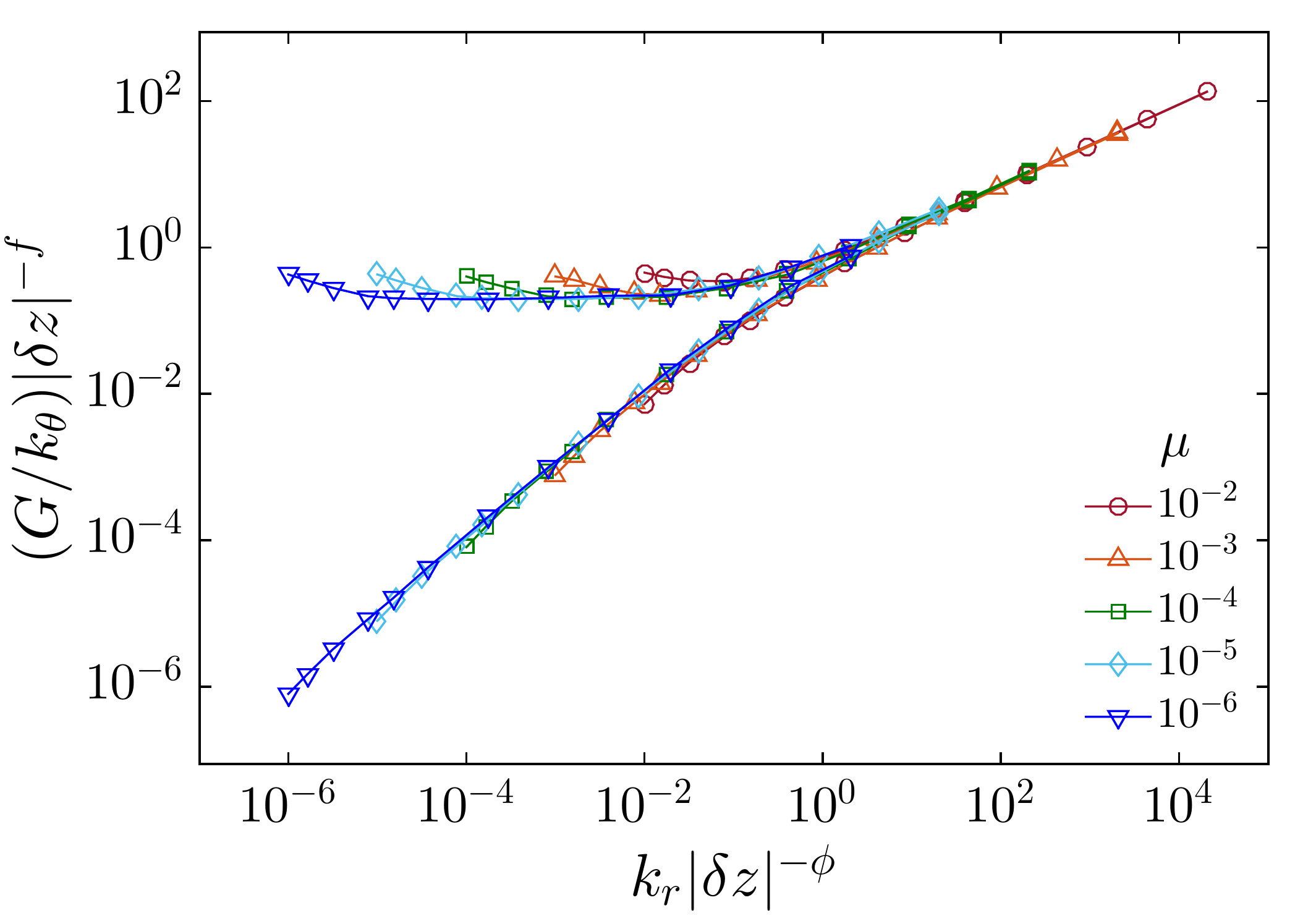}
\caption{Scaling collapse of the shear modulus $G$; here the scaling exponents $f=1.25$ and $\phi = 2.25$ give the best collapse.}
\label{scaling_collapse_G}
\end{figure}

In Fig.~\ref{scaling_collapse_G} we show a scaling plot for $G$; here $G/k_\theta$ is rescaled by $|\delta z|^f$ and plotted against the ratio $k_r/|\delta z|^{\phi}$ where $\delta z\!\equiv\! z\!-\! z_c$ denotes the distance to the critical coordination $z_c\!\equiv\!4$. The best collapse is found using the exponents $f\!=\!1.25$ and $\phi\!=\!2.25$. In what follows we will argue that the mean field exponents are $f\!=\!1$ and $\phi\!=\!2$, and discuss the discrepancy we find with the mean field predictions. 

To understand the scaling behavior of the shear modulus as seen in Figs.~\ref{shear_modulus_vs_coordination} and \ref{scaling_collapse_G}, in the next Sections we explain the occurrence of an isostatic point at $z_c\!=\!4$ in our model, and consider the limit $\mu\!\to\!0$ separately in the hyperstatic $z>z_c$ and the hypostatic $z<z_c$ regimes. 

\section{The angle-preserving isostatic point of 2D planar networks}
\label{isostatic_point}
In this Section we extend the Maxwell-Calladine linear-algebraic constructions for the rigidity of frames of struts \cite{maxwell1864calculation,calladine1978buckminster} to 3-body (bond-bending) geometries, in order to establish that the angle-preserving isostatic point of 2D planar networks is $z_c=4$.

The emergence of an isostatic point in 2D planar networks becomes apparent when the limit $\mu\!\to\!0$ is considered, as seen in Fig.~\ref{shear_modulus_vs_coordination}. In the hypostatic regime $z\!\le\! z_c$ the limit $\mu\!\to\!0$ can be obtained by fixing $k_r$ and sending $k_\theta\!\to\!\infty$. Under these circumstances the bond-bending interactions can be treated as \emph{geometric constraints}; a displacement field $\uv_\ell$ on the network's nodes will leave the angles $\theta_{ijk}$ invariant (to leading order in $\uv_\ell$) if it satisfies
\begin{equation}\label{foo11}
\frac{\partial \theta_{ijk}}{\partial\xv_\ell}\cdot\uv_\ell = 0\,,
\end{equation}
for every angle $\theta_{ijk}$, where here and in what follows repeated node indices are understood to be summed over. An example of the field $\frac{\partial \theta_{ijk}}{\partial\xv_\ell}$ is shown in Fig.~\ref{fig1}b.

It is useful to define the linear operator
\begin{equation}\label{foo12}
{\cal Q}_{ijk,\ell} \equiv \frac{\partial \theta_{ijk}}{\partial\xv_\ell}\,,
\end{equation}
which takes vectors from the space of the nodes' coordinates to the space defined by the entire set of angles. The number of rows ${\cal Q}$ features is equal to the total number of angles $Nz$, whereas the number of columns is $2N$, each represented a spatial coordinate of a node. Nontrivial solutions to Eq.~\ref{foo11} are expected to exist if the rank of the operator ${\cal Q}$ is \emph{smaller} than the number of degrees of freedom $2N$ available to the displacement field $\uv$. 

We next argue that the rank of ${\cal Q}$ becomes exactly equal to $2N$ at $z\!=\!4$; to this aim we define $z_i$ to be the number of edges connected to the $i\th$ node, and we note that the number of angles that surround each node is equal to $z_i$. A displacement field that preserves $z_i\!-\!1$ angles that surround a single node must also preserve the $z_i\th$ angle as well, meaning that a single angle for each node can be selected, and the row of ${\cal Q}$ corresponding to that particular angle can be eliminated (it will be expressible as a linear combination of other rows). This amounts to eliminating $N$ rows out of the $Nz$ rows of ${\cal Q}$. 

We finally note that each face of our planar network is a polygon with $m$ edges, built from $m$ angles; a displacement field that preserves $m\!-\!1$ angles of a face must also preserve the $m\th$ angle, further reducing the number of independent rows of ${\cal Q}$ by the number ${\cal F}$ of faces of the network. The latter is related to the number of nodes $N$ and the number of edges $\frac{1}{2}Nz$ via Euler's formula for a planar graph embedded on a torus
\begin{equation}
{\cal F} = \sFrac{1}{2}Nz - N\,.
\end{equation}

The total number of independent rows left after the eliminations described above is thus
\begin{equation}
\mbox{rank}(Q) = Nz - N - \big(\sFrac{1}{2}Nz - N\big) = \sFrac{1}{2}Nz\,,
\end{equation}
which becomes equal to the dimension of configuration space $2N$ when the coordination reaches $z_c\!=\!4$. In these considerations we neglect corrections of order $N^{-1}$ that arise from collective translations or deformations of space~\cite{PhysRevE.90.022138}. 

We conclude that in our two-dimensional disordered networks nontrivial displacement fields that preserve the entire set of angles will exist if $z\!<\!4$, whereas when $z\!>\!4$ no such displacements exist. Interestingly, the critical coordination that separates these two regimes coincides with the Maxwell threshold $z_c\!=\!2\dbar$ of networks of rigid struts \cite{maxwell1864calculation,calladine1978buckminster} discussed intensively in the jamming literature, see e.g.~\cite{van_hecke,liu_review}.

Having established that an underlying isostatic point exists at $z\!=\!4$, we next turn to discussing the scaling behavior of elastic moduli as the critical coordination is approached from above and from below.

\section{The hyperstatic regime $z\!>\! z_c$}
\label{hyperstatic}

It is illuminating to study the elasticity of our model in the hyperstatic regime by fixing $k_\theta\!=\!1$ and sending $k_r\!\to\!0$, resulting in $\mu\!\to\!0$ (see black circles in Fig.~\ref{shear_modulus_vs_coordination}a). These circumstances correspond to eliminating the second sum on the RHS of Eq.~(\ref{potential_energy}) for the potential energy, leaving us with
\begin{equation}
U_{k_r\!=\!0} = \frac{k_\theta \bar{\ell}^2}{2}\sum_{\left< i,j,k\right>} \Delta{ \theta_{ijk} }^2 \,.
\end{equation}
In the remainder of this Section we show how in this limit elastic moduli expressions can be simply expressed via the geometry of the network, and provide scaling arguments to moduli's dependence on coordination for $z\!>\!z_c$.

\subsection{Elastic moduli in the hyperstatic regime}
\label{hyperstatic_moduli}

The microscopic expression for the athermal shear modulus $G$ for systems governed by a potential energy $U$ reads \cite{lutsko}
\begin{equation}\label{foo13}
G = \frac{1}{V}\left(\frac{\partial^2U}{\partial\gamma^2} - \frac{\partial^2U}{\partial\gamma\partial\xv_\ell}\cdot{\cal M}^{-1}_{\ell m}\cdot\frac{\partial^2U}{\partial\xv_m\partial\gamma}\right)\,,
\end{equation}
where 
\begin{equation}
{\cal M}\equiv\frac{\partial^2U}{\partial\xv\partial\xv}
\end{equation}
is known as the \emph{dynamical matrix}. Recall that our systems are assumed to have $U\!=\!0$ (i.e.~all of the angles reside exactly at their rest-angles) before any deformation is imposed, leading to a simple form for the dynamical matrix; it reads
\begin{equation}
{\cal M}_{\ell m} = k_\theta\bar{\ell}^2\sum_{\left< i,j,k\right>}\frac{\partial\theta_{ijk}}{\partial\xv_\ell}\frac{\partial\theta_{ijk}}{\partial\xv_m}\,.
\end{equation}
Working in units such that $\bar{\ell}\!=\!1$ and $k_\theta\!=\!1$, employing Eq.~\ref{foo12} and adopting bra-ket and matrix notations, the dynamical matrix takes the form
\begin{equation}\label{foo14}
{\cal M} = {\cal Q}^T{\cal Q}\,,
\end{equation}
where here and in what follows we assume that the redundent rows of ${\cal Q}$ have been eliminated as described in Sect.~\ref{isostatic_point}.  

\begin{figure*}[t]
\includegraphics[width= 0.9\textwidth]{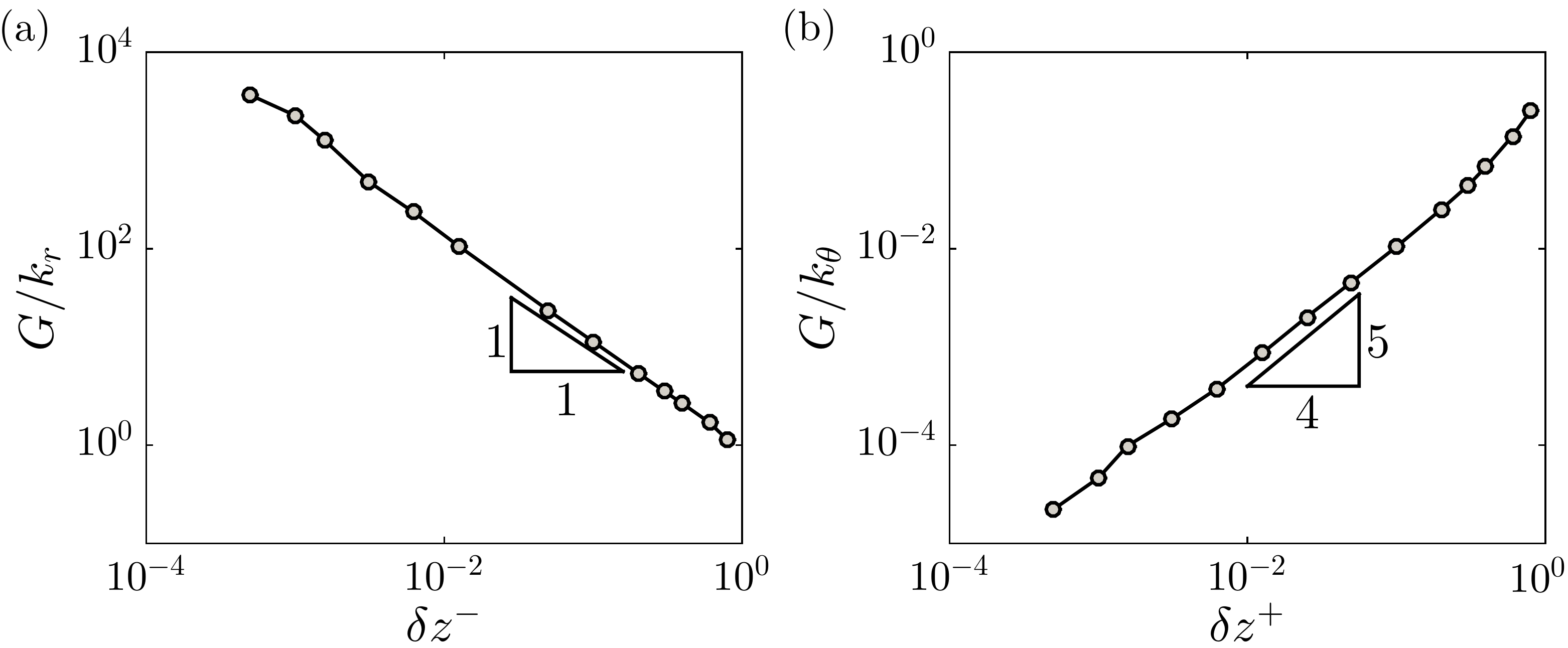}
\caption{The shear modulus $G$ in the $\mu\!=\!0$ limit. (a) In the hypostatic regime, where $z<z_c$, we plot $G$ rescaled by the stiffness $k_r$, as a function of the distance to the critical coordination $\delta z^-$. (b) In the hyperstatic regime where $z>z_c$, we plot $G$ rescaled by the stiffness $k_\theta$, as a function of the distance to the critical coordination $\delta z^+$.}
\label{limitsShear}
\end{figure*}

We next see that under the circumstances of fixing $k_\theta\!=\!1$ and sending $k_r$ to zero,
\begin{equation}
\frac{\partial^2U}{\partial\xv_\ell\partial\gamma} = \sum_{\left< i,j,k\right>}\frac{\partial \theta_{ijk}}{\partial \xv_\ell}\frac{\partial \theta_{ijk}}{\partial\gamma}\,,
\end{equation}
which can be expressed using our bra-ket notation as
\begin{equation}\label{foo15}
\ket{\partial^2_{\xv,\gamma}U} = {\cal Q}^T\ket{\partial_\gamma \theta}\,,
\end{equation}
where we denoted $\partial_\gamma\!\equiv\!\partial/\partial\gamma$ and $\partial^2_{\xv,\gamma}\!\equiv\!\partial^2/\partial\gamma\partial\xv$. We note that the space of angles $\theta_{ijk}$ is assumed here to only consist of those angles that were not eliminated from the corresponding rows of ${\cal Q}$. 

Combining now Eqs.~(\ref{foo13}),(\ref{foo14}) and (\ref{foo15}), we arrive at a simple expression for the shear modulus:
\begin{equation}
G = \frac{\bra{\partial_\gamma\theta}{\cal I} - {\cal Q}({\cal Q}^T{\cal Q})^{-1}{\cal Q}^T\ket{\partial_\gamma \theta}}{V}\,.
\end{equation}
A similar expression was first put forward in \cite{matthieu_thesis} for the case of random networks of relaxed Hookean springs. 

\subsection{Scaling argument for hyperstatic moduli}
\label{hyperstatic_argument}

We follow a similar line of argumentation as presented in \cite{matthieu_thesis} for elastic moduli of random networks of relaxed Hookean springs. From general considerations (see e.g.~\cite{Lerner2018} for a detailed discussion) it can be shown that for $z\!>\! z_c$
\begin{equation}
{\cal I} - {\cal Q}({\cal Q}^T{\cal Q})^{-1}{\cal Q}^T = \sum_\ell\ket{\phi_\ell}\bra{\phi_\ell}\,,
\end{equation}
where $\ket{\phi_\ell}$ are the zero-modes of the operator ${\cal Q}{\cal Q}^T$, namely they satisfy
\begin{equation}
{\cal Q}^T\ket{\phi_\ell} = \zerovector\,.
\end{equation}
These objects are akin to the so-called \emph{states of self stress} studied intensively in the context of the jamming transition \cite{wouter_epl_2009, ellenbroek_rigidity_prl_2015, sussman_sss, Lerner2018} and the physics of topological metamaterials \cite{Kane2014,Paulose2015}. Simple counting arguments \cite{matthieu_thesis,wouter_epl_2009,sussman_sss} suggest that in a system of size $N$ with coordination $z$ there are $N(z\!-\! z_c)$ orthonormal modes $\ket{\phi_\ell}$. Assuming these modes are extended and random objects one expects $\braket{\partial_\gamma\theta}{\phi_\ell}\!\sim\!{\cal O}(1)$, and therefore we predict
\begin{equation}\label{foo20}
G = \frac{1}{V}\sum_\ell\braket{\partial_\gamma\theta}{\phi_\ell}^2 \sim z-z_c\,,
\end{equation}
in the hyperstatic regime, and in the limits $z\!\to\! z_c^{+}$ and $\mu\!\to\!0$. This prediction is also in agreement with Effective Medium calculations, see e.g.~\cite{Wyart_2010,eric_boson_peak_emt}; it suggests that the scaling exponent $f\!=\!1$, c.f.~Fig.~\ref{scaling_collapse_G}. 

In Fig.~\ref{limitsShear}b we show our measurements of the shear modulus for $k_\theta\!=\!1$ and $k_r\!=\!0$. We do not find perfect agreement with the theoretical prediction. However, this disagreement with the scaling argument and mean field theory seems not to be unique to our system in which bending interactions dominate; the shear modulus of randomly diluted hyperstatic spring networks in 2D shows a similar super linear scaling with $\delta z$ too, as demonstrated in previous works \cite{PhysRevLett.107.158303, PhysRevLett.120.148004}. We speculate \cite{private_communication2} that dimensionality may be playing a role. Uncovering the origin of these observed disagreements is left for future work. 

\begin{figure}[!ht]
\includegraphics[width= 0.48\textwidth]{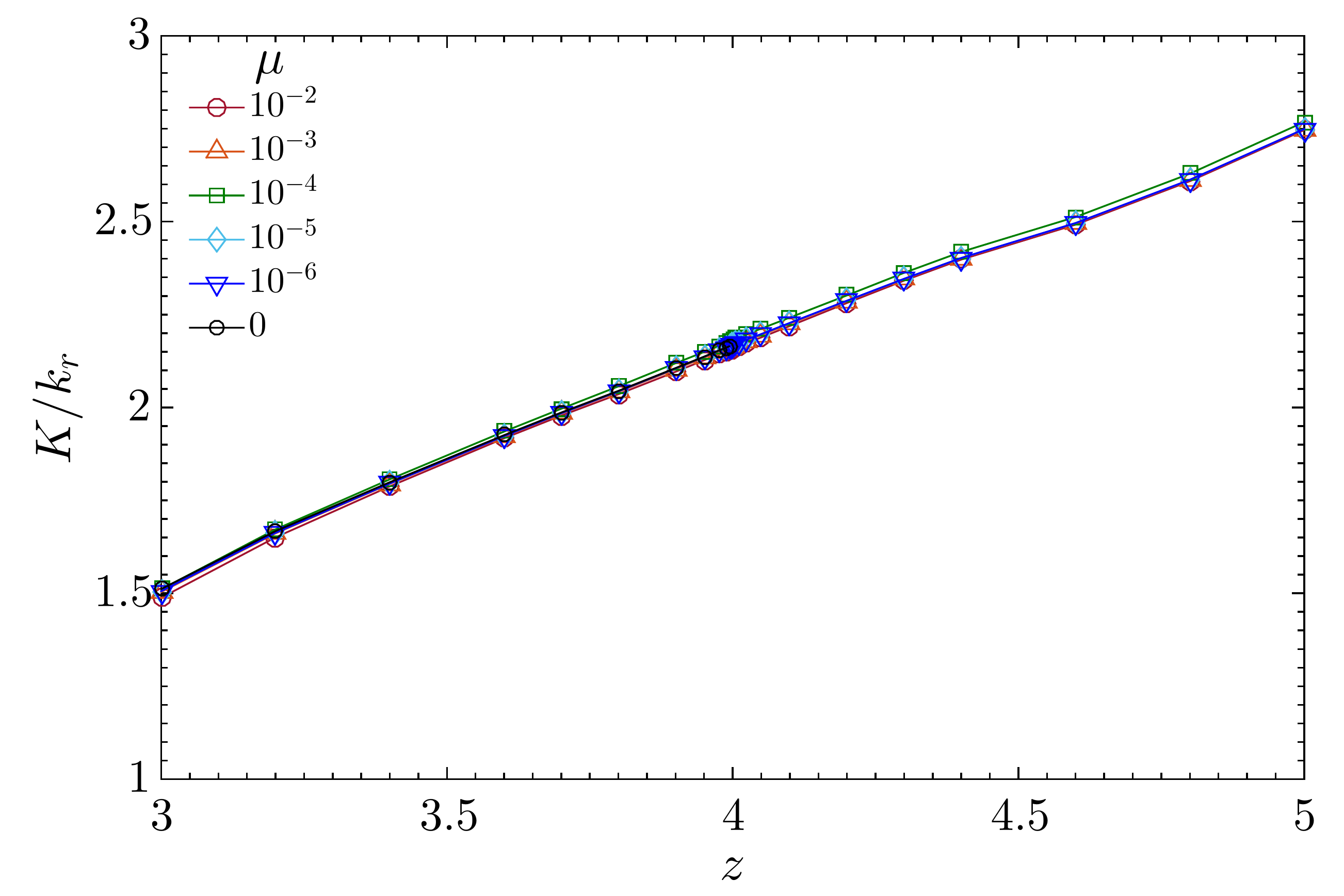}
\caption{The bulk modulus $K$ divided by the stiffness $k_r$, as a function of the coordination $z$ for various values of the stiffness ratio $\mu$.}
\label{bulkmod}
\end{figure}

As for the bulk modulus, we note that isotropic expansions leave the angles $\theta_{ijk}$ unchanged. As a consequence, we expect the bulk modulus to trivially scale with $k_r$ throughout the entire coordination range, and to trivially grow with increasing coordination. In Fig.~\ref{bulkmod} we show that this is indeed the case. 

\section{The hypostatic regime $z\!<\! z_c$}
\label{hypostatic}

To study the elasticity of our model in the hypostatic regime, we fix $k_r$ and send $k_\theta\!\to\!\infty$, resulting in $\mu\!\to\!0$. In this route of taking the $\mu\!\to\!0$ limit the bond-bending interactions can be treated as \textbf{geometrical} constraints, as mentioned in Sect.~\ref{isostatic_point}. 

In what follows we assume $z\!<\!z_c$, we fix $k_r$ and send $k_\theta\!\to\!\infty$, resulting in $\mu\!\to\!0$. We aim at incorporating the geometric constraints of fixed angles while deriving microscopic expressions for elastic moduli; a similar derivation was recently presented in \cite{robbies_pre} in the context of the nonlinear mechanics of biopolymeric fibrous materials. In what follows we will consider the angles $\theta_{ijk}$ as being entirely fixed; the potential energy of the material then reduces to
\begin{equation}\label{only_radial_U}
U_{k_\theta\!=\!0} =\frac{k_r}{2}\sum_{\left< i,j\right>} \Delta{ r_{ij} }^2\,,
\end{equation}
i.e.~we only consider the central-force part of the energy. We aim at taking the derivative of the energy with respect to the imposed deformation, under two sets of constraints: $(i)$ the system must remain in mechanical equilibrium along the deformation, and $(ii)$ the geometric constraints of fixed angles must be satisfied by the total displacements of the network's nodes. 

In order to satisfy condition $(i)$, we introduce Lagrange multipliers $\tau_{ijk}$ that correspond to clamping torques associated with each fixed angle $\theta_{ijk}$; in terms of these torques and the potential energy given by Eq.~(\ref{only_radial_U}), the net force experienced by the $i\th$ node must vanish, namely
\begin{equation}\label{foo05}
\fv_m = \sum_{\left< i,j,k\right>}\tau_{ijk}\frac{\partial \theta_{ijk}}{\partial\xv_m} - \frac{\partial U}{\partial\xv_m} = 0\,.
\end{equation}
Condition $(i)$ is satisfied by demanding that not only is the net force zero, but it also does not change under imposed deformations, namely
\begin{equation}\label{foo03}
\frac{d \fv_m}{d\gamma} = \frac{\partial\fv_m}{\partial\gamma} + \frac{\partial \fv_m}{\partial\xv_n}\cdot\yv_n +\sum_{\left< i,j,k\right>}\frac{\partial \fv_m}{\partial\tau_{ijk}}\frac{d\tau_{ijk}}{d\gamma} = 0\,,
\end{equation}
where the \emph{nonaffine displacements} $\yv$ are additional displacements that the nodes must perform on top of the imposed deformation, in order to satisfy the mechanical equilibrium constraints, and leave the angles invariant under the external deformation. The latter requirement can be expressed as a constraint equation for each angle $\theta_{ijk}$ that reads
\begin{equation}\label{foo02}
\frac{d\theta_{ijk}}{d\gamma} = \frac{\partial\theta_{ijk}}{\partial\gamma} + \frac{\partial\theta_{ijk}}{\partial \xv_m}\cdot\yv_m = 0\,.
\end{equation}
Setting $k_r\!=\!1$, and defining the linear operator \cite{calladine1978buckminster}
\begin{equation}\label{foo21}
{\cal S}_{ij,k} \equiv \frac{\partial r_{ij}}{\partial\xv_k}\,,
\end{equation}
we show in Appendix~\ref{derivations} that Eq.~(\ref{foo03}) and (\ref{foo02}) can be incorporated into a single relation as 
\begin{equation}\label{foo04}
\left( \begin{array}{cc}{\cal S}^T{\cal S}&-{\cal Q}^T\\-{\cal Q}&0\end{array}\right)
\left( \begin{array}{c}\ket{\yv}\\\ket{\dot{\tau}}\end{array}\right) = 
\left( \begin{array}{c}-{\cal S}^T\ket{\partial_\gamma r} \\ \ket{\partial_\gamma \theta}\end{array}\right)\,,
\end{equation}
where we have used the notations $\partial_\gamma r_{ij}\!\equiv\!\partial r_{ij}/\partial\gamma$ and $\partial_\gamma \theta_{ijk}\!\equiv\!\partial \theta_{ijk}/\partial\gamma$. Eq.~(\ref{foo04}) forms a closed linear system that can be inverted in favor of the nonaffine displacement field $\yv$ and the torque variations $\dot{\tau}$. 

\subsection{Elastic moduli in the hypostatic regime}

We are now in the position to derive expressions for elastic moduli, in the limit $\mu\!\to\!0$ obtained by taking $k_\theta\!\to\!\infty$ and $k_r\!=\!1$; the second derivatives of the energy with respect to deformation reads
\begin{eqnarray}\label{foo06}
\frac{d^2U}{d\gamma^2} & = & \frac{\partial^2 U}{\partial\gamma^2} + 2\frac{\partial^2U}{\partial\gamma\partial\xv_k}\cdot\yv_k \nonumber \\
&& + \frac{\partial ^2U}{\partial\xv_k\partial\xv_m}:\yv_k\yv_m + \frac{\partial U}{\partial\xv_k}\cdot\frac{d\yv_k}{d\gamma}\,.
\end{eqnarray}
Assuming again an unstressed material, namely that all torques $\tau_{ijk}\!=\!0$ vanish then following Eq.~(\ref{foo05})
\begin{equation}\label{foo16}
\frac{\partial U}{\partial\xv_k} =\zerovector\,.
\end{equation}
Notice further that for the simple form of the potential Eq.~(\ref{only_radial_U}), and by setting $k_r\!=\!1$, one finds 
\begin{equation}\label{foo17}
\frac{\partial^2U}{\partial\xv_k\partial\xv_\ell} = \sum_{\left< i,j\right>}\frac{\partial r_{ij}}{\partial\xv_k}\frac{\partial r_{ij}}{\partial\xv_\ell}\  \leftrightarrow \ {\cal S}^T{\cal S}\,,
\end{equation}
and
\begin{equation}\label{foo18}
\frac{\partial^2U}{\partial\xv_k\partial\gamma} = \sum_{\left< i,j\right>}\frac{\partial r_{ij}}{\partial\xv_k}\frac{\partial r_{ij}}{\partial\gamma}\  \leftrightarrow\  {\cal S}^T\ket{\partial_\gamma r}\,.
\end{equation}
Combining Eqs.~(\ref{foo16}), (\ref{foo17}) and (\ref{foo18}) with Eq.~(\ref{foo06}), we arrive at an expression for elastic moduli in the hypostatic regime $z\!<\!z_c$ in the limit $\mu\!\to\!0$:
\begin{equation}\label{foo09}
G = \frac{\braket{\partial_\gamma r}{\partial_\gamma r} + 2\bra{\partial_\gamma r}{\cal S}\ket{\yv} + \bra{\yv}{\cal S}^T{\cal S}\ket{\yv}}{V}\,.
\end{equation}

\subsection{Scaling arguments for hypostatic moduli}
\label{hypostatic_argument}
Examining Eq.~(\ref{foo09}) it is clear that if the characteristic scale of nonaffine displacements $y\!\equiv\!\sqrt{\braket{\yv}{\yv}/N}$ diverges as $z\!\to\! z_c$, one expects to observe scaling laws of elastic moduli with respect to $z\!-\! z_c$. To understand the behavior of the nonaffine velocities, we rearrange Eq.~(\ref{foo04}) in favor of $\yv$ as
\begin{eqnarray}
\ket{\yv} & =&  ({\cal S}^T{\cal S})^{-1} {\cal S}^T\ket{\partial_\gamma r} 
-{({\cal S}^T{\cal S})}^{-1}{\cal Q}^T  \big( {\cal Q}{({\cal S}^T{\cal S})}^{-1}{\cal Q}^T\big)^{-1}  \nonumber\\ & &\cdot \big[{\cal Q}{({\cal S}^T{\cal S})}^{-1}{\cal S}^T\ket{\partial_\gamma r} - \ket{\partial_\gamma\theta} \big]\,.
\end{eqnarray}
We now perform a mean-field approximation, and consider a potential energy that consists of Hookean springs of unit stiffness that connect each node to its absolute initial position $\xv^{(0)}$, namely
\begin{equation}
U_{\mbox{\tiny mf}} = \sum_i |\xv_i - \xv_i^{(0)}|^2\,.
\end{equation}
In this case the dynamical matrix $\partial^2U_{\mbox{\tiny mf}}/\partial\xv\partial\xv$ reduces to the identity tensor ${\cal I}$ (instead of ${\cal S}^T{\cal S}$, as given by Eq.~(\ref{foo17})), and the nonaffine displacements assume the simple form
\begin{equation}\label{foo19}
\ket{\yv} ={\cal S}^T\ket{\partial_\gamma r} + {\cal Q}^T ({\cal Q}{\cal Q}^T)^{-1}({\cal Q}{\cal S}^T\ket{\partial_\gamma r} - \ket{\partial_\gamma\theta})\,.
\end{equation}
Recall that when $z\!=\! z_c$ a single solution $\ket{\phi}$ to the equation ${\cal Q}^T\ket{\phi}\!=\!\zerovector$ appears. We therefore expect the operator ${\cal Q}{\cal Q}^T$ to posses lower and lower frequency modes as $z$ approaches $z_c$, that, in turn, give rise to diverging nonaffine displacements. This is akin to the behavior observed for the operator ${\cal S}{\cal S}^T$ in floppy networks \cite{during2013phonon}.

\fboxsep=0mm
\begin{figure}[b]
\begin{center}
 \raisebox{100pt}{(a)} \fbox{\includegraphics[height=100pt]{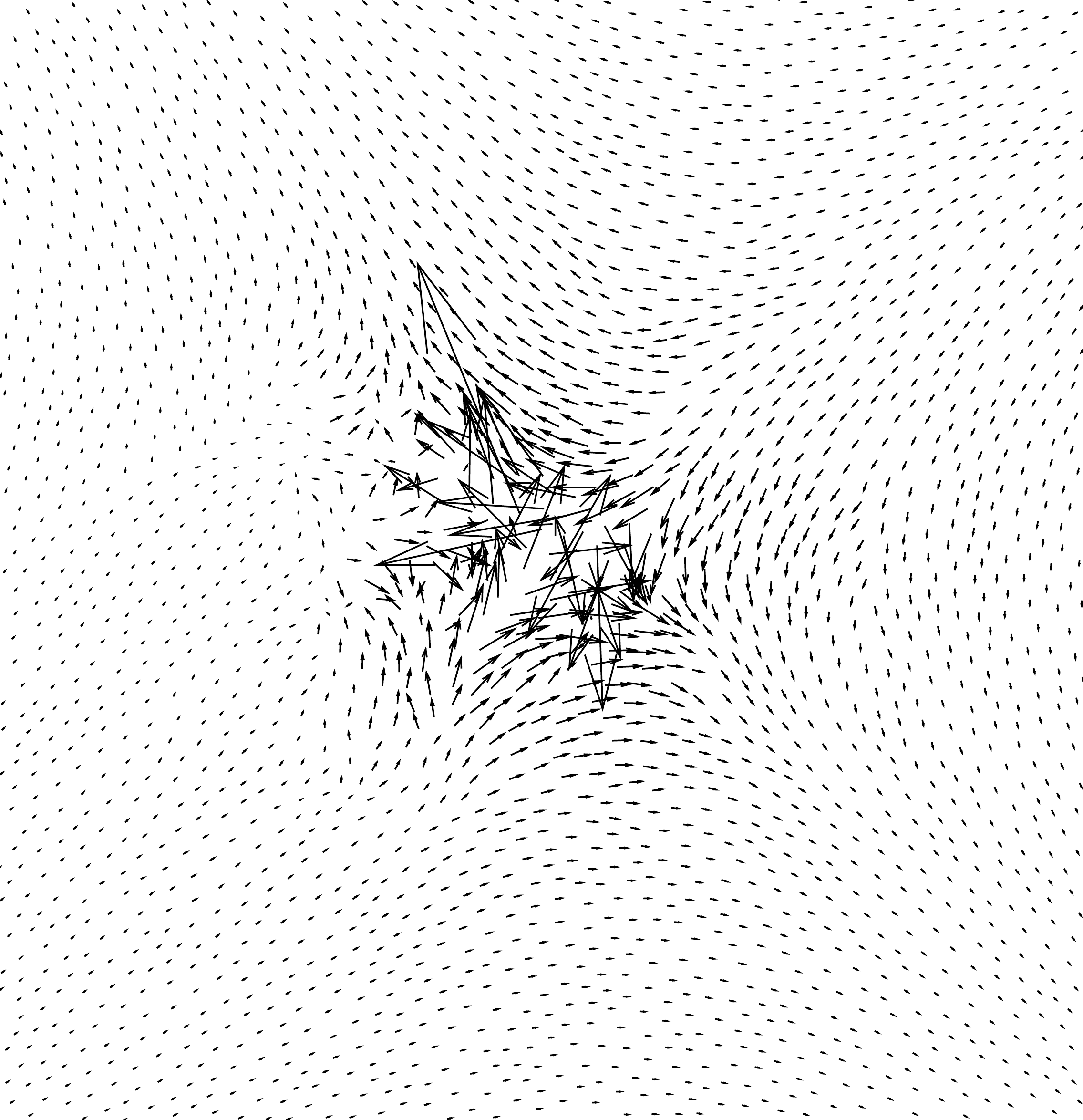}} \quad  \raisebox{100pt}{(b)}  \fbox{\includegraphics[height=100pt]{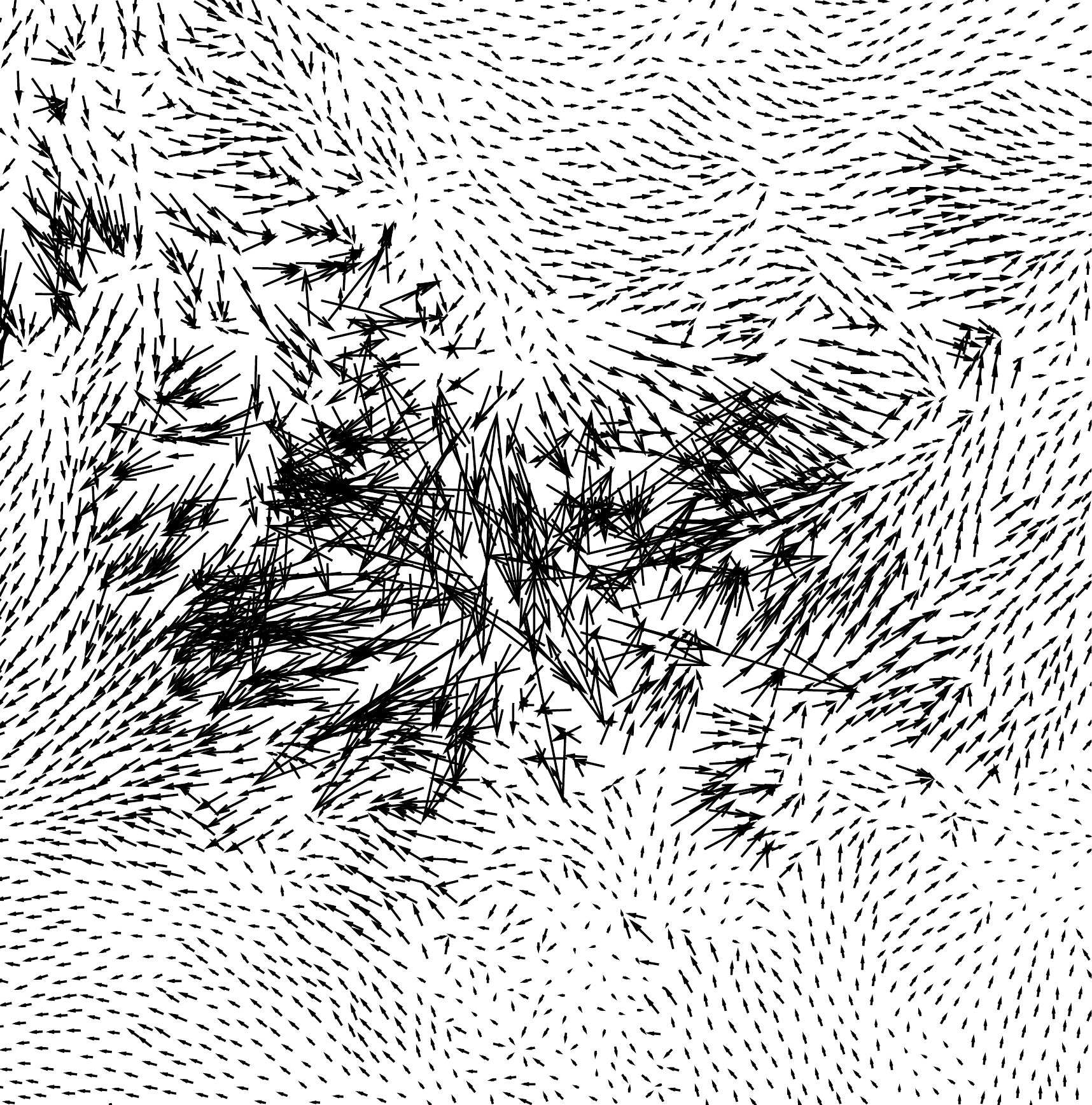}}
\caption{Displacement response field  $\delta \xv$ of two networks of $N = 1\ 000\ 000$ with coordination (a) $z = 4.4$ and (b) $z =4.05$, zoomed in on the core of the localized force perturbation.}
\label{field}
\end{center}
\end{figure}

We next note that the first term in the right hand side (RHS) of Eq.~(\ref{foo19}) is regular, and in Appendix~\ref{negligible} we argue that the first term in the brackets on the RHS of Eq.~(\ref{foo19}) (the term involving ${\cal Q}$) can be neglected compared to the second term as $z\!\to\! z_c$. Therefore, as $z\!\to\!z_c$, the characteristic scale of nonaffine displacements follows
\begin{equation}
y^2 \equiv \braket{\yv}{\yv}/N \sim \bra{\partial_\gamma\theta}\big({\cal Q}{\cal Q}^T\big)^{-1}\ket{\partial_\gamma\theta}\,.
\end{equation} 
Assuming that the operator ${\cal Q}$ is random, it has been shown in \cite{Beltukov2015} that the spectrum of concatenations of the form ${\cal Q}{\cal Q}^T$ should depend on the dimensions of the operator ${\cal Q}$; in particular, one expects the density of states (i.e.~the distribution of the square root of the eigenvalues) of ${\cal Q}{\cal Q}^T$ to follow the Marchenko-Pastur distribution \cite{Beltukov2015}, which in the small frequency and $z\!\to\!z_c$ limits takes the form
\begin{equation}
D(\omega) \sim \frac{\sqrt{\omega^2 - \omega_\star^2}}{\omega\omega_\star^2}\,,
\end{equation}
with $\omega_\star\!\sim\! z_c\!-\! z$. Assuming next that the eigenmodes of ${\cal Q}{\cal Q}^T$ are random, extended objects (similarly to the arguments made before Eq.~(\ref{foo20})), we estimate
\begin{equation}
y^2 \sim \int\frac{D(\omega)}{\omega^2}d\omega \sim \int_{\omega_\star}\frac{d\omega}{\omega^2} \sim \frac{1}{\omega_\star} \sim \frac{1}{z_c - z}\,.
\end{equation}

Finally, following Eq.~(\ref{foo09}) we expect that to leading order in $y$, $G\!\sim\! y^2$. We therefore conclude this Section with the prediction
\begin{equation}\label{foo22}
G \sim \frac{1}{z_c -z}\,,
\end{equation}
in the hypostatic regime, in the limits $z\!\to\! z_c^{-}$ and $\mu\!\to\!0$.

In Fig.~\ref{limitsShear}a we plot $G$ vs.~$\delta z$ for our hypostatic systems, and find excellent agreement with the theoretical prediction Eq.~(\ref{foo22}). We note that similar results were shown for an elastic system subjected to radial constraints in \cite{robbies_pre}. Eq.~(\ref{foo22}) implies the scaling relation $f\!-\!\phi=-1$; using the mean-field exponent $f\!=\!1$ (see Sect.~\ref{hyperstatic_argument}), one expects $\phi\!=\!2$. However, the best collapse in Fig.~\ref{scaling_collapse_G} is achieved using $f\!=\!1.25$ and $\phi\!=\!2.25$, consistent with our predicted scaling relation, and with the direct measurement in the hyperstatic regime shown in Fig.~\ref{limitsShear}. 


\begin{figure*}[t]
\begin{center}
\includegraphics[width= 0.9\textwidth]{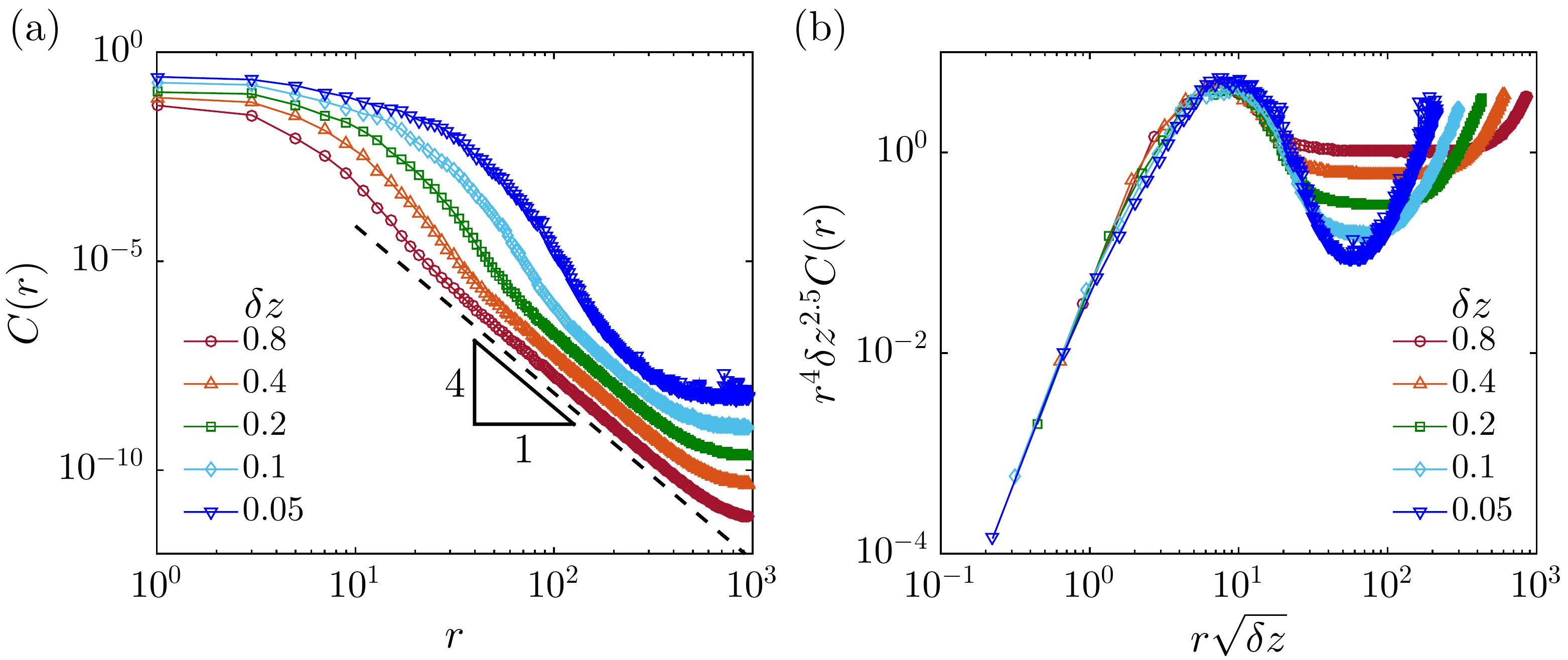}
\caption{Amplitude of the response fields to a local perturbation (see definition in Eq.~(\ref{decay_def})), as a function of the distance $r$ to the local imposed forcing. The signal is averaged over 50 responses to different pertubations, calculated in networks of $N = 1\ 000\ 000$ nodes and with varying coordinations in the hyperstatic regime as indicated by the legend. The far fields show a $r^{-4}$ decay, in agreement with similar analyses shown in \cite{breakdown,Lerner2018}.}
\label{decay_collapse_fig}
\end{center}
\end{figure*}

\section{Diverging lengthscale in the hyperstatic regime}
\label{lengthscale}

In Sections~\ref{hyperstatic} and \ref{hypostatic} we rationalize the scaling of the shear modulus in the limit $\mu\!\to\!0$ both in the hyperstatic and hypostatic regimes, respectively. Building on the discussions held in both of these Sections, we now make a scaling argument that predicts a diverging lengthscale in the hyperstatic regime, that is expected to follow
\begin{equation}
\ell_c \sim \frac{1}{\sqrt{z-z_c}}\,.
\end{equation}

The argument is made as follows; consider the vibrational spectra of our bending dominated disordered networks with $\mu\!=\!0$ and $z$ larger than but close to the critical coordination $z_c$. As shown in Sect.~\ref{hyperstatic_moduli}, for $\mu\!=\!0$ the dynamical matrix assumes the form
\begin{equation}
{\cal M} = {\cal Q}^T{\cal Q}\,.
\end{equation}
We note that, apart from possible zero modes, the spectra of the concatenations ${\cal Q}{\cal Q}^T$ and of ${\cal Q}^T{\cal Q}$ are identical \cite{Beltukov2015}. We therefore expect the occurrence of a plateau of disordered vibrational modes above the frequency scale $\omega_\star\!\sim\! z\!-\! z_c$ \cite{Beltukov2015}, as discussed in Sect.~\ref{hypostatic_argument}.

On the other hand, our system's potential energy is invariant to global translations, and therefore Goldstone modes are expected to be present at small frequencies. In particular, the frequency $\tilde{\omega}_{\mbox{\tiny min}}$ of the longest-wavelength Goldstone mode depends on the system size $L$ and the shear modulus $G$ as 
\begin{equation}
\tilde{\omega}_{\mbox{\tiny min}} \sim \sqrt{G}/L\,.
\end{equation}

Consider now the response to a localized force perturbation in a random bending-dominated network of coordination $z$ and linear size $L$; if $\tilde{\omega}_{\mbox{\tiny min}}\!\ll\!\omega_\star$, we expect the far field of the response to the local perturbation to exhibit a continuum-elastic-like structure, whilst if $\tilde{\omega}_{\mbox{\tiny min}}\!\gg\!\omega_\star$ no continuum-elastic-like response is expected since low-frequency disordered modes will overwhelm the response. Since the two frequency scales $\tilde{\omega}_{\mbox{\tiny min}}$ and $\omega_\star$ become comparable when $L$ is of the order of $1/\sqrt{z\!-\! z_c}$, we expect to see a signature of a diverging length $\ell_c\!\sim\!1/\sqrt{z\!-\! z_c}$ in the spatial structure of the response to point perturbations. 

We stress that both in floppy (hypostatic) \cite{during2013phonon} and in hyperstatic random networks \cite{breakdown} the length $\ell_c$ was observed, as well in several other previous works. Ref.~\cite{Lerner2018} provides a comprehensive summary of additional observations of the length $\ell_c$ in the existing literature.

To test our argument, we select randomly an angle $\theta_{ijk}$, and impose a localized force perturbation of the form
\begin{equation}
\fv_m = \frac{\partial\theta_{ijk}}{\partial\xv_m}\,.
\end{equation}
An example of this force can be seen in Fig.~\ref{fig1}b. 
The linear displacement response to the force $\fv_m$ reads
\begin{equation}
\delta \xv_\ell = {\cal M}^{-1}_{\ell m}\cdot\fv_m = {\cal M}^{-1}_{\ell m}\cdot\frac{\partial \theta_{ijk}}{\partial\xv_m}\,.
\end{equation}
We denote any two angles in the system $\theta_1$ and $\theta_2$, and by $r_{12}$ the distance between the nodes associated with these angles. In Fig.~\ref{field} an example of the response field is visualized for two different coordinations. The growing length scale is clearly showing. We next define
\begin{equation}\label{decay_def}
C(r) = \bigg<\frac{\partial\theta_1}{\partial\xv_m}\cdot{\cal M}^{-1}_{m\ell}\cdot\frac{\partial\theta_2}{\partial\xv_\ell}\bigg>_{r_{12}}\,,
\end{equation}
where $\langle\rangle_{r_{12}}$ denotes the average over all pairs of angles $\theta_1,\theta_2$ separated by a distance $r_{12}$. Fig.~\ref{decay_collapse_fig} shows the results of our numerical calculations of $C(r)$: in panel (a) we show the mean of $50$ different force perturbations. In panel (b) we plot the same data shown in panel (a), this time scaled by $r^4\delta z^{2.5}$, and plotted against $r\sqrt{\delta z}\!\sim\! r/\ell_c$, clearly revealing that the lengthscale governing the transition to a continuum-like response is $\ell_c\!\sim\!1/\sqrt{z\!-\! z_c}$, consistent with our scaling argument.

\section{Auxeticity}
\label{auxeticity}
In this penultimate Section we discuss the auxetic behavior of our model material. Auxeticity is quantified via the Poisson's ratio $\nu$, defined in 2D as
\begin{equation}
\nu = \frac{K-G}{K+G}\,.
\end{equation}
Materials possessing $\nu\!=\!1/2$ are known as \emph{incompressible}, whereas materials with $\nu\!<\!0$ are termed \emph{auxetic}; the latter are nongeneric, and as such draw attention in the field of architected metamaterials \cite{Bouwsta,Yang2004,ReidE1384}. 

\begin{figure}[!ht]
\includegraphics[width= 0.47\textwidth]{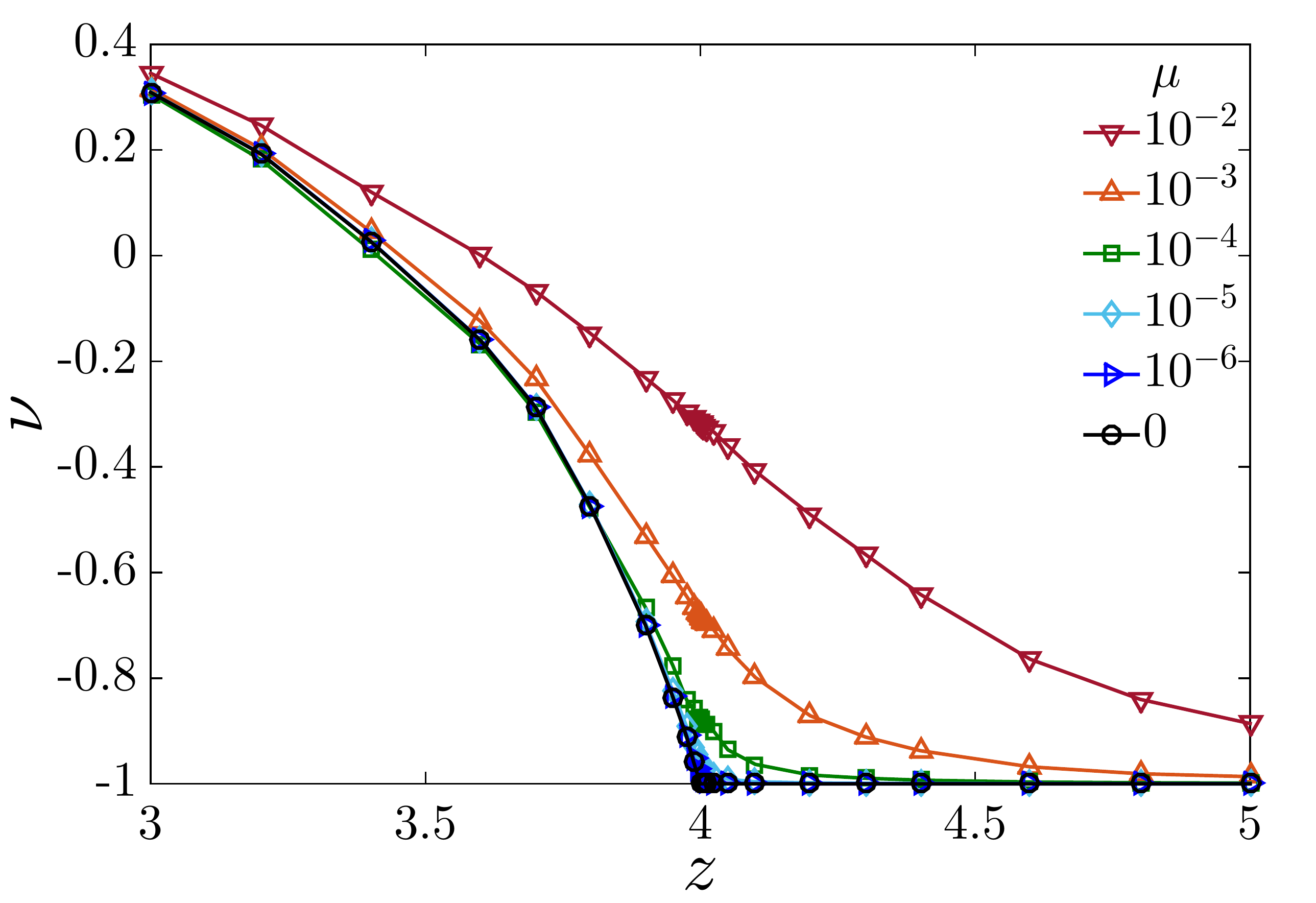}
\caption{The Poisson's ratio $\nu$ as a function of the coordination $z$ for various values of the ratio of stiffnesses $\mu$.}
\label{auxetic_fig}
\end{figure}

In Fig.~\ref{auxetic_fig} we plot the Poisson's ratio $\nu$ averaged over 20 realizations as a function of the coordination, for various values of the stiffness ratio $\mu$. We find that our system becomes auxetic in the entire range of $\mu$ explored ($\mu\!\le\!10^{-2}$), for $z$ roughly larger than 3.5. As expected from the scaling of $G$ and $K$ discussed previously, in the limit $\mu\!\to\!0$ we find a transition to perfect auxeticity $\mu\!=\!-1$ at the isostatic point $z_c\!=\!4$. From the scaling behavior of $G$ discussed in Sect.~\ref{hypostatic} we expect $\nu\!+\!1 \sim z_c\!-\! z$ as $z\!\to\! z_c$, as supported by our data shown in Fig.~\ref{auxetic_fig}.

\section{Summary and Discussion}
\label{discussion}
In this work we explored the elastic properties of a model system that represents a material whose mechanics is dominated by bond-bending interactions, rather than the commonly-studied steric or radial interactions. Our study was motivated by recent intriguing experimental work by Schall and coworkers \cite{schall_inspiration}, who fabricated colloidal superstructures using critical Casimir forces \cite{C7SM00599G}, in which bond-bending interactions between the constituent patchy colloidal particles were shown to be much stiffer than radial ones. 

In our model system we observe numerically and rationalize theoretically the existence of an angle-preserving isostatic point $z_c\!=\!4$ that governs an underlying jamming transition observed when the ratio of stretching to bending stiffnesses of the interactions vanishes. The jamming behavior we observe as the coordination is made to approach the critical coordination from below is reminiscent of the strain stiffening transition observed in fibrous biological materials \cite{rens2016nonlinear,jansen2018,robbies_pre,Merkel6560}. 

Our theoretical arguments for the scaling of elastic moduli in the limits $\mu\!\to\!0$, $z\!\to\! z_c^+$ (the hyperstatic regime) and $z\!\to\! z_c^{-}$ (the hypostatic regime) follow closely two previous theoretical approaches to the jamming \cite{matthieu_thesis} and the strain-stiffening \cite{robbies_pre} transitions in random networks, respectively. In the hyperstatic regime, the arguments put forward by Wyart \cite{matthieu_thesis} using the operator ${\cal S}$ (see Eq.~(\ref{foo21})) --- that represents the constraints associated with radial interactions --- were applied here using the operator ${\cal Q}$ that represents constraints associated with conserving the angles of our networks. In the hypostatic regime, our argumentation echos the framework and reasoning presented in \cite{robbies_pre}, where again the role of the operator ${\cal S}$ in \cite{robbies_pre} is played by the operator ${\cal Q}$ in the present work. The applicability of these approaches establishes their generality, and highlights the key physical ingredient governing the mechanics near jamming and stiffening transitions -- the interplay between interactions-induced constraints and configurational degrees of freedom. 

\acknowledgements
We warmly thank G.~D\"uring and C.~Coulais for fruitful discussions. E.~L.~acknowledges support from the Netherlands Organisation for Scientific Research (NWO) (Vidi grant no.~680-47-554/3259). R.R. and  E.L.~acknowledge support from the Delta Institute for Theoretical Physics (D-ITP consortium), a program of NWO that is funded by the Dutch Ministry of Education, Culture and Science (OCW).

Author contribution statement: both R.R.~and E.L.~designed and carried out the research, and wrote the article.

\appendix

\section{Network generation protocol}
\label{network_generation_protocol}
The protocol for generating homogeneous networks consists of a two-step process. In the first step, a packing of soft disks is compressed up to a certain pressure. Using a bidisperse distribution of radii we prevent crystallization and ensures a disordered realization. A contact is assigned to each overlapping disc. We then distillate the contact network to obtain a highly coordinated network of edges and nodes, with an average of between 5 and 6 edges per node.

The second step is a biased bond dilution protocol that aims at maintaining homogeneity of the network, which is typically lost in a truely random dilution. The network with $N$ nodes and a set of edges $\cal E$.  An edge $e_{ij}\!\in\!{\cal E}$ connects between a pair of neighboring nodes $i,j$, with ranks $z_i$ and $z_j$ respectively. For every edge $e_{ij}$ we define the sum of ranks $s_{ij}\! =\! z_i\! +\! z_j$ and the absolute difference $d_{ij}\!=\!\left | z_i\! -\! z_j \right |$. The edges with largest value for $s_{ij}$ and than smallest value of $d_{ij}$, will remove fluctuations in coordination of the network when removed. This min-max protocol typically produces a large set of edges that are eligible to be removed. An additional selection condition is checked to suppress the creation of sharp angles between edges. 
For each edge in the selected set the remaining angle after bond removal is calculated. The bond with the minimal remaining angle is now removed from network.  
This process is repeated until the desired connectivity is achieved. 

As an example, in Fig.~\ref{fig1} (a), one finds the network diluted to a connectivity of $3.95$.  
To reiterate, this methods serves particularly well in generating a network with low connectivity fluctuations.

\section{Equations of motion for nonaffine displacements and clamping torques variations}
\label{derivations}
In this Appendix we show how Eq.~\ref{foo04} is obtained from Eqs.~\ref{foo05}, \ref{foo03} and \ref{foo02}.
The requirement of maintaining force balance under an imposed deformation is given by Eq.~\ref{foo03}; it reads
\begin{equation} \label{boo01}
\frac{d \fv_m}{d\gamma} = \frac{\partial\fv_m}{\partial\gamma} + \frac{\partial \fv_m}{\partial\xv_n}\cdot\yv_n +\sum_{\left< i,j,k\right>}\frac{\partial \fv_m}{\partial\tau_{ijk}}\frac{d\tau_{ijk}}{d\gamma} = 0\,,
\end{equation}
Where $\fv_m$ is the sum of all forces acting on node $m$, which is given by equation \ref{foo02},
\begin{equation} \label{boo02}
\fv_m = \sum_{\left< i,j,k\right>}\tau_{ijk}\frac{\partial \theta_{ijk}}{\partial\xv_m} - \frac{\partial U}{\partial\xv_m} = 0\,.
\end{equation}
Each term in Eq. \ref{boo01} contains a partial derivative of the net force $\fv$, worked out below using the definition given by Eq.~\ref{boo02}, namely
\begin{eqnarray}
\frac{\partial\fv_m}{\partial\gamma} & = & - \frac{\partial^2 U}{\partial \gamma  \partial\xv_m}\,, \nonumber\\ 
\frac{\partial\fv_m}{\partial\xv} & = & - \frac{\partial^2 U}{\partial \xv \partial\xv_m} \,, \nonumber\\ 
\frac{\partial\fv_m}{\partial\tau_{ijk}} & = & \sum_{\left< i,j,k\right>}\frac{\partial \theta_{ijk}}{\partial\xv_m} \,, \nonumber
\end{eqnarray}
where terms containing $\tau$ are dropped because in the undeformed state the torque forces are assumed to be identically zero. 

To simply further consider the specified potential energy given by 
\begin{equation}
U_{k_\theta\!=\!0} =\frac{k_r}{2}\sum_{\left< i,j\right>} \Delta{ r_{ij} }^2\,.
\end{equation}
In undeformed state all pairwise distances in the network are at their rest length, so $\Delta r_{ij}\! =\! 0$ for each $\left< i,j\right>$. 
Setting $k_r\!=\!1$, the second order derivative $\partial^2_{\xv,\xv}  U $ reduces to
\begin{equation}
\frac{ \partial ^2 U}{\partial \xv \partial \xv} 
 							= \sum_{\left< i,j\right>} \frac{ \partial r_{ij}}{\partial \xv} \frac{ \partial r_{ij}}{\partial \xv}  \,,
\end{equation}
and the mixed derivative  $\partial^2_{\gamma,\xv}  U$ to  
\begin{equation}
\frac{ \partial ^2 U}{\partial \gamma \partial \xv} =   \sum_{\left< i,j\right>} \frac{ \partial r_{ij}}{\partial \gamma} \frac{ \partial r_{ij}}{\partial \xv}  \,,
\end{equation}
Finally we employ operator and braket notations, where 
\begin{equation}
{\cal Q}_{ijk,\ell} \equiv \frac{\partial \theta_{ijk}}{\partial\xv_\ell}\,, \quad \mbox{and}\quad  {\cal S}_{ij,k} \equiv \frac{\partial r_{ij}}{\partial\xv_k}\,.
\end{equation}
Combining all of the above, Eq.~\ref{boo01} can now be compactly written as  
\begin{equation} \label{boo03}
\ket{ \dot  \fv}  = -\partial_\gamma {\cal S} -( {\cal S}^T {\cal S} ) \ket{ \yv} +{\cal Q} \ket{ \dot \tau} = 0\,,
\end{equation}
and similarly the invariance of the angles, given by Eq.~\ref{foo02}, as
\begin{equation}\label{boo04}
\ket{\dot \theta} = \ket{\partial_\gamma \theta} + {\cal Q}\ket{\yv} = 0\,,
\end{equation}
with $\partial_\gamma\!\equiv\!\partial/\partial\gamma$.
Combining equations \ref{boo03} and \ref{boo03} into one set of equations, we obtain
\begin{equation}\label{boo05}
\left( \begin{array}{cc}{\cal S}^T{\cal S}&-{\cal Q}^T\\-{\cal Q}&0\end{array}\right)
\left( \begin{array}{c}\ket{\yv}\\\ket{\dot{\tau}}\end{array}\right) = 
\left( \begin{array}{c}-{\cal S}^T\ket{\partial_\gamma r} \\ \ket{\partial_\gamma \theta}\end{array}\right)\,,
\end{equation}
as given by Eq.~(\ref{foo04}) in the main text. 

\vspace{0.3cm}

\section{Leading order term in Eq.~\ref{foo19}}
\label{negligible}
In Eq.~\ref{foo19} the second term on the RHS contains the sum of two terms and reads ${\cal Q}^T ({\cal Q}{\cal Q}^T)^{-1}({\cal Q}{\cal S}^T\ket{\partial_\gamma r} - \ket{\partial_\gamma\theta})$. 
Here we argue that the vector $\ket{b} = \left ( {\cal Q}{\cal S}^T\ket{\partial_\gamma r} - \ket{\partial_\gamma\theta} \right )$ can be approximated as $\ket{b} \sim - \ket{\partial_\gamma\theta}$ near the critical point. The reason is that the first of these two terms features an additional factor {\cal Q}; since this term is expected to be regular when contracted with ${\cal Q}^T ({\cal Q}{\cal Q}^T)^{-1}$ (and see similar discussion in \cite{robbies_pre}), it can be neglected in our scaling analysis. 

%

\end{document}